 \documentclass{emulateapj}

\usepackage{amsmath}
\usepackage{amssymb}
\usepackage{graphicx}
\usepackage{epsfig}
\usepackage{color}
\usepackage{natbib}

\shorttitle{Gamma-Ray Bursts in Circumstellar Shells}
\shortauthors{Mesler et al.}

\begin{document}

\title{Gamma-Ray Bursts in Circumstellar Shells: A Possible Explanation for Flares}







\author{R. A. Mesler\altaffilmark{1}, Daniel J. Whalen\altaffilmark{2,3}, Nicole M. Lloyd-Ronning\altaffilmark{3}, Chris L. Fryer\altaffilmark{3} and Y. M. Pihlstr\"om\altaffilmark{1}}

\altaffiltext{1}{Department of Physics and Astronomy, University of New Mexico, Albuquerque, NM  87131}
\altaffiltext{2}{McWilliams Fellow, Department of Physics, Carnegie Mellon University, Pittsburgh, PA 15213}
\altaffiltext{3}{Los Alamos National Laboratory, Los Alamos, NM 87545}
 

\begin{abstract}

It is now generally accepted that long-duration gamma ray bursts (GRBs) are due to the collapse of massive rotating stars.  The precise collapse process itself, however, is not yet fully understood.  Strong winds, outbursts, and intense ionizing UV radiation from single stars or strongly interacting binaries are expected to destroy the molecular cloud cores that give birth to them and create highly complex circumburst environments for the explosion.  Such environments might imprint features on GRB light curves that uniquely identify the nature of the progenitor and its collapse.  We have performed numerical simulations of realistic environments for a variety of long-duration GRB progenitors with ZEUS-MP, and have developed an analytical method for calculating GRB light curves in these profiles.  Though a full, three-dimensional, relativistic magnetohydrodynamical computational model is required to precisely describe the light curve from a GRB in complex environments, our method can provide a qualitative understanding of these phenomena.  We find that, in the context of the standard afterglow model, massive shells around GRBs produce strong signatures in their light curves, and that this can distinguish them from those occurring in uniform media or steady winds. These features can constrain the mass of the shell and the properties of the wind before and after the ejection.  Moreover, the interaction of the GRB with the circumburst shell is seen to produce features that are consistent with observed X-ray flares that are often attributed to delayed energy injection by the central engine.  Our algorithm for computing light curves is also applicable to GRBs in a variety of environments such as those in high-redshift cosmological halos or protogalaxies, both of which will soon be targets of future surveys such as \textit{JANUS} or \textit{Lobster}.

\end{abstract}




\section{Introduction}

It has long been believed, and pre-explosion progenitors have proven for nearly a dozen cases, that type Ib/c and II supernovae are produced in the collapse of massive stars.  Although we have yet to observe a pre-explosion gamma-ray burst (GRB) progenitor, the evidence that many long-duration GRBs are also produced via massive star collapse has grown since their discovery in 1973 \citep[for reviews, see][]{woosleyBloom06,fryerEA07b}.  The engine at the heart of these long-duration GRBs is believed to be powered either by a rapidly-accreting black hole or a rapidly-spinning magnetar \citep{woosley93,woosleyBloom06,barkovKomissarov11}, but the exact progenitor of these engines is still unknown. More than a dozen scenarios have been proposed in which a massive star or tightly-coupled binary system can collapse to form such disks \citep{fryerWoosley98,fryerEA99b, fryerEA07b,zhangFryer01}.

The simplest progenitor is the classic collapsar, a massive star that sheds its hydrogen envelope by strong winds and violent mass eruptions (akin to luminous blue variable outbursts) prior to the collapse of its core \citep{woosley93}.  Additionally, a host of massive star models exist, a set of which invoke binary merger events.  Like the violent mass ejection in massive stars, these merger events eject shells of material into the immediate surroundings before the GRB outburst.  Both single star and binary mass ejections can occur just a few thousand years prior to stellar collapse.  In the case of the helium merger model, we expect the merger-driven mass ejection to occur less than a few years before the launch of the GRB jet.  The jet must plow through this shell as it is producing the gamma-ray emission we observe.

The structure of the circumburst media (CSM) should imprint signatures on their afterglows that identify the mode of collapse \citep{meszaros02, woosley11}, providing a mechanism that can be used to better understand the progenitors of these cosmic explosions.  Analytical models of relativistic jets in both uniform circumburst densities and free-streaming wind profiles have yielded light curves that are in reasonable agreement with observations \citep{bergerEA00, yostEA03, curranEA11, priceEA02}.  Recently, collisions of the jet with more complicated structures have been examined, such as wind-termination shocks \citep{ramirez-RuizEA01, ramirez-RuizEA05, daiLu02, nakarGranot07}, shocks due to collisions between stellar winds from nearby stars. \citep{mimicaGiannios11}, clumps \citep{ramirez-RuizEA05}, and magnetic shocks \citep{yostEA03}.  \citet{ramirez-RuizEA01, ramirez-RuizEA05} and \citet{daiLu02} found that sharp features in circumburst densities due to clumps and shocks create discernible features in GRB light curves, although \citet{nakarGranot07}, who consider the dynamics of the reverse shock and assume that the jet remains relativistic after encountering a jump in density, do not.  \citet{mimicaGiannios11} found gradual shallowing in the light curve when the jet passes through a shock formed when stellar winds from the progenitor and a neighbor collide.

Flares are often observed in GRB lightcurves within the first few thousand seconds, notably in GRB 081029 \citep{nardiniEA11}, GRB 071112C \citep{huangEA12a}, GRB 050502B \citep{falconeEA06}, GRB 060607A \citep{ziaeepourEA08}, and GRB 050421 \citep{godetEA06}.  It is generally accepted that these flares are a product of late-time injections of energy into the system by the GRB's central engine, but this is not the only possible scenario.  When the jet from a GRB encounters a shell produced by a stellar progenitor, it experiences an abrupt increase in the medium density. GRB light curves are much more sensitive to a decrease in medium density than to an increase \citep{yostEA03}, meaning that an abrupt increase in density of the order of $\sim$ a few will leave almost no discernible imprint on the observed light curve.  In contrast, our hydrodynamical models can produce sudden enhancements in the density of five orders of magnitude or more at the trailing edge of the shell (Fig.\ \ref{figure1_windBubbles}), which is sufficient to produce bright flares despite the weak dependence of the light curve on an increase in density.     

Winds, outbursts and ionizing UV radiation from GRB progenitors all disperse the molecular cloud cores that created them and create far more complex ambient morphologies for the jet than those considered in afterglow studies to date \citep[but note][]{fryerEA06,whalenEA08b}.  In this paper, we model such environments for a variety of GRB scenarios with ZEUS-MP.  Our 1D models span a wide variety of winds and outbursts with complex gas chemistry and cooling that capture the true structure of the circumburst medium.  We have also developed a semi-analytical approach based on previous work by \citet{panaitescuKumar00}, \citet{huangEA99}, and \citet{peer12} for computing GRB light curves in any general density profile, not just the uniform media, winds and simple density jumps in previous studies.  We apply this new method to compute light curves for relativistic jets propagating through circumburst shells, and examine the imprint of these shells on the light curves in order to determine if they constrain the mode of collapse.  

In $\S \, 2$ we discuss the ZEUS-MP code, how it is used to simulate the environments of long-duration GRBs, and our grid of wind models.  We review the results of our shell ejection calculations in $\S \, 3$.  In $\S \, 4$ we describe how our new analytical models of GRB jets are applied to hydrodynamical profiles from these simulations to compute light curves for a variety of energies.  We calculate GRB light curves in our grid of shell profiles, which correspond to a variety of collapse scenarios, and determine if specific light curve features constrain the nature of the progenitor.  In $\S \, 5$ we conclude.


\section{Numerical Method}

\subsection{ZEUS-MP}

ZEUS-MP is a massively-parallel astrophysical hydrodynamics code that solves nonequilibrium H and He gas chemistry and photon-conserving ionizing UV radiation transport together with Eulerian fluid dynamics in a self-consistent manner \citep{whalenNorman06, whalenNorman08b, whalenNorman08a}.  The hydrodynamics equations are

\vspace{0.1in}
\begin{eqnarray}
\frac{\partial \rho}{\partial t}  & = & - \nabla \: \cdotp \; (\rho {\bf v})  \\
\frac{\partial \rho v_{i}}{\partial t}  & = & - \nabla \: \cdotp \; (\rho v_{i} 
{\bf v}) \: - \: \nabla p \: - \: \rho \nabla \Phi \: - \: \nabla \cdotp {\bf Q}    \\ 
\frac{\partial e}{\partial t}  & = & - \nabla \: \cdotp \; (e {\bf v}) \: - \: p\nabla \: 
\cdotp \: {\bf v} \: - \: \bf{Q} : \nabla  {\bf v}, 
\end{eqnarray} \vspace{0.05in} 

\noindent where $\rho$, $e$, and the $v_{i}$ are the gas density, internal energy density, and velocity of each zone and $p = (\gamma-1)\, e$ and {\bf{Q}} are the gas pressure and the von Neumann-Richtmeyer artificial viscosity tensor.  We evolve mass fractions for H, H$^{+}$, He, He$^{+}$, He$^{2+}$, H$^{-}$, H$^{+}_{2}$, H$_{2}$, and e$^{-}$ with nine additional continuity equations (the species are assumed to have the same velocities) and the nonequilibrium rate equations of \citet{anninosEA97} 

\vspace{0.05in}
\begin{equation}
\frac{\partial \rho_{i}}{\partial t} = - \nabla \: \cdotp \; (\rho {\bf v}) 
+ \sum_{j}\sum_{k} {\beta}_{jk}(T){\rho}_{j}{\rho}_{k} + \sum_{j} {\kappa}
_{j}{\rho}_{j}, \label{eqn:network}\vspace{0.05in}
\end{equation}

where ${\beta}_{jk}$ is the rate coefficient for the reaction between species j and k that creates (+) or destroys (-) species i, and the ${\kappa}_{j}$ are the radiative reaction rates.  Microphysical cooling and heating are calculated with operator-split isochoric updates to the gas energy density that are performed every time the reaction network is solved: 

\vspace{0.05in}
\begin{equation}
{\dot{e}}_{gas} = \Gamma - \Lambda. \label{eqn: egas}
\vspace{0.05in}  
\end{equation}

Here, $\Gamma$ is the photoionization heating rate for all species over all photon energies and $\Lambda$ is the sum of all cooling rates.  We include collisional excitation and ionization cooling by H and He, recombinational cooling, H$_2$ cooling, and bremsstrahlung cooling, with our nonequilibrium reaction network providing the species mass fractions needed to accurately 
calculate these collisional cooling processes.  We also calculate fine structure cooling due to C, O, N, Si and Fe using the \citet{dalgarnoMcCray72} cooling curves, generalized to arbitrary elemental abundances.  We exclude cooling by dust.  

Fluid flow, gas heating and cooling, and H and He chemistry can occur on highly disparate timescales whose relative magnitudes can widely vary throughout the course of a calculation.  The many chemical reaction timescales can be consolidated into a single chemistry time step defined as 

\begin{equation}
t_{chem} = 0.1 \, \displaystyle\frac{n_{e}}{{\dot{n}}_{e}}, \label{eqn: t_chem}
\end{equation}

which is formulated to ensure that the fastest reaction operating at any place or time on the grid determines the maximum time by which the reaction network may be accurately advanced.  The timescale on which the gas heats or cools is given by

\vspace{0.1in}
\begin{equation}
t_{h/c} = 0.1 \, \displaystyle\frac{e_{gas}}{{\dot{e}}_{gas}}. \label{eqn: t_hc}
\end{equation}

To evolve each physical process on its respective timescale without restricting the entire algorithm to the shortest one, we subcycle the reaction network and energy equation over the time step on which we update the hydrodynamics equations.  We first compute the minimum $t_{chem}$ and $t_{h/c}$ for the entire grid and then perform consecutive updates of species mass fractions and gas energy densities over the smaller of these two times until the lesser of $t_{h/c}$ and $t_{CFL}$ is reached, where $t_{CFL}$ is the Courant time.  At this point a full update of the hydrodynamics equations is performed and the cycle repeats.

The prefactor of 0.1 in equations \ref{eqn: t_chem} and \ref{eqn: t_hc} guarantees that mass fractions never change by more than 10\% when the reaction network is solved and that gas energies do not change by more than 10\% over a time step.  This prevents catastrophic runaway cooling in gas at high densities and metallicities by either H$_2$ or fine-structure cooling, like those in dense shells swept up by strong winds or mass ejections.

\subsection{Problem Setup}

To model mass ejections in ZEUS-MP, we treat stellar winds and outbursts as time-dependent inflows at the inner boundary of a 1D spherical grid with 32,000 zones.  The gas is assigned H and He mass fractions of 0.76 and 0.24, respectively, and a metallicity $Z =$ 0.1 $Z_{\odot}$.  The mesh extends from 10$^{-4}$ pc to 0.3 pc with outflow conditions on the outer boundary. 
The inflow is imposed at the inner boundary in the form of a time-varying density and velocity: 

\vspace{0.05in}
\begin{equation}
\rho \, = \, \frac{\dot{m}}{4 \pi {r_{ib}}^2 v_{w}}, \vspace{0.05in}
\end{equation}

where $r_{ib}$ is the radius of the inner boundary and $v_{w}$ is the wind velocity.  Outbursts are modeled by increasing $\dot{m}$ and lowering $v_{w}$.  Because stellar winds clear out gas from the vicinity of the star prior to any outbursts, we initialize the grid with a free-streaming density and velocity profile

\vspace{0.05in}
\begin{equation}
\rho(r) \, = \, \frac{\dot{m}}{4 \pi r^2 v_{w}}, \vspace{0.05in}
\end{equation}

where the wind velocity is assumed to be constant.  The temperatures of the initial density profile is set to 100 K.  We launch the outburst at the beginning of the simulation.  The grid is domain decomposed into 8 tiles, with 4000 mesh zones per tile and one tile per processor.  

We neglect the effect of ionizing radiation from the star on the structure of the dense shell.  This treatment is approximate, given that the progenitor illuminates the flow over its entire lifetime and that its luminosity evolves over this period.  However, the heat deposited in the wind by photoionizations is small in comparison to its bulk kinetic energy and is unlikely to alter the properties of the flow in the proximity of the GRB. 

\subsection{Grid of Shell Models}

For collapsar and He mergers we consider 3 mass loss rates $\dot{m}_w = 10^{-6}, 10^{-5}$, and $10^{-4}$ M$_{\odot}$/yr and outbursts $\dot{m}_b = 10^{-2}$ M$_{\odot}$/yr lasting for 10 yr and 100 yr that correspond to total shell masses of 0.1 and 1.0 M$_{\odot}$, respectively.  We take the velocities of the fast wind and slow shell to be 2000 km/s and 200 km/s, respectively, and in each model use the given $\dot{m}_w$ to initialize the density across the entire grid, assuming that the star sheds mass at the same rate before and after the outburst.  In general, larger shell masses are expected for He mergers but both kinds of progenitors can exhibit light, moderate and heavy winds, so there is some degeneracy across our grid of models.  We also consider He-He mergers by simulating the loss of the common envelope with a single massive outburst $\dot{m}_b$ = 10 M$_{\odot}$/yr that lasts for one year and has a velocity of 200 km/s.  

\subsection{Wind Bubble Test}

Before calculating circumburst environments for a GRB progenitor, we first model the bubble blown by its fast wind during the life of the star. This bubble is the primary circumstellar structure formed by the star and grows to radii of 20 - 30 pc before outbursts alter its interior on scales of a few tenths of a parsec late in the life of the star.  Since the gamma-ray spectra and afterglow of the GRB are primarily governed by the interaction of the jet with its surroundings out to $\sim$ 
0.1 pc and 10 pc, respectively, this primary shell does not affect the observational signature of the GRB at early times.  However, we simulate this bubble first to verify that for reasonable choices of ambient density it is indeed at least 10 pc from the star when it dies and does not immediately impact the afterglow.  Also, the effect of local density and radiative cooling on the structure and kinematics of shells in general is more easily seen with this first bubble than the more complicated structures created by interactions between slow outbursts and fast winds just before the death of the star.

\begin{figure*}
\plottwo{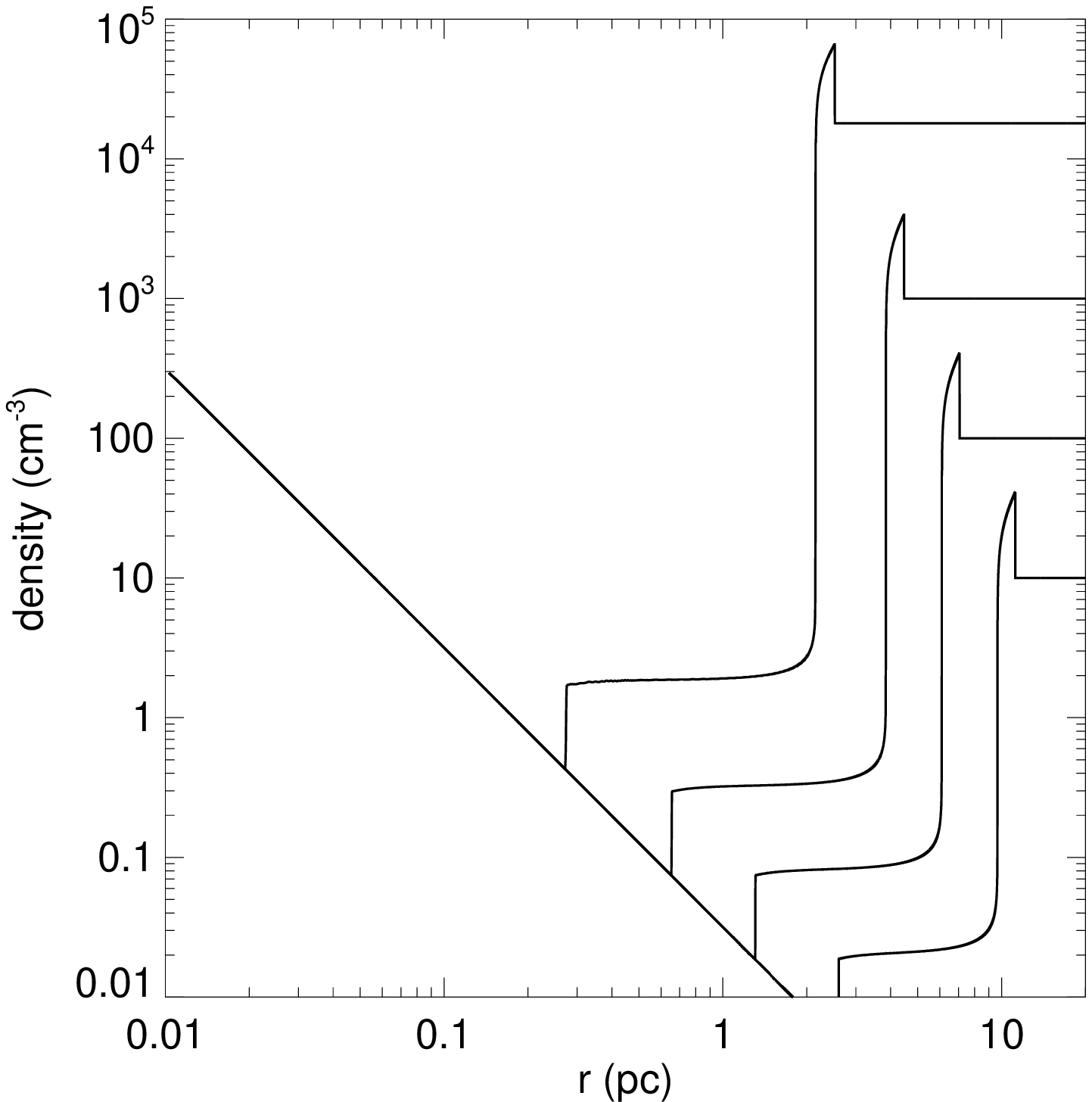}{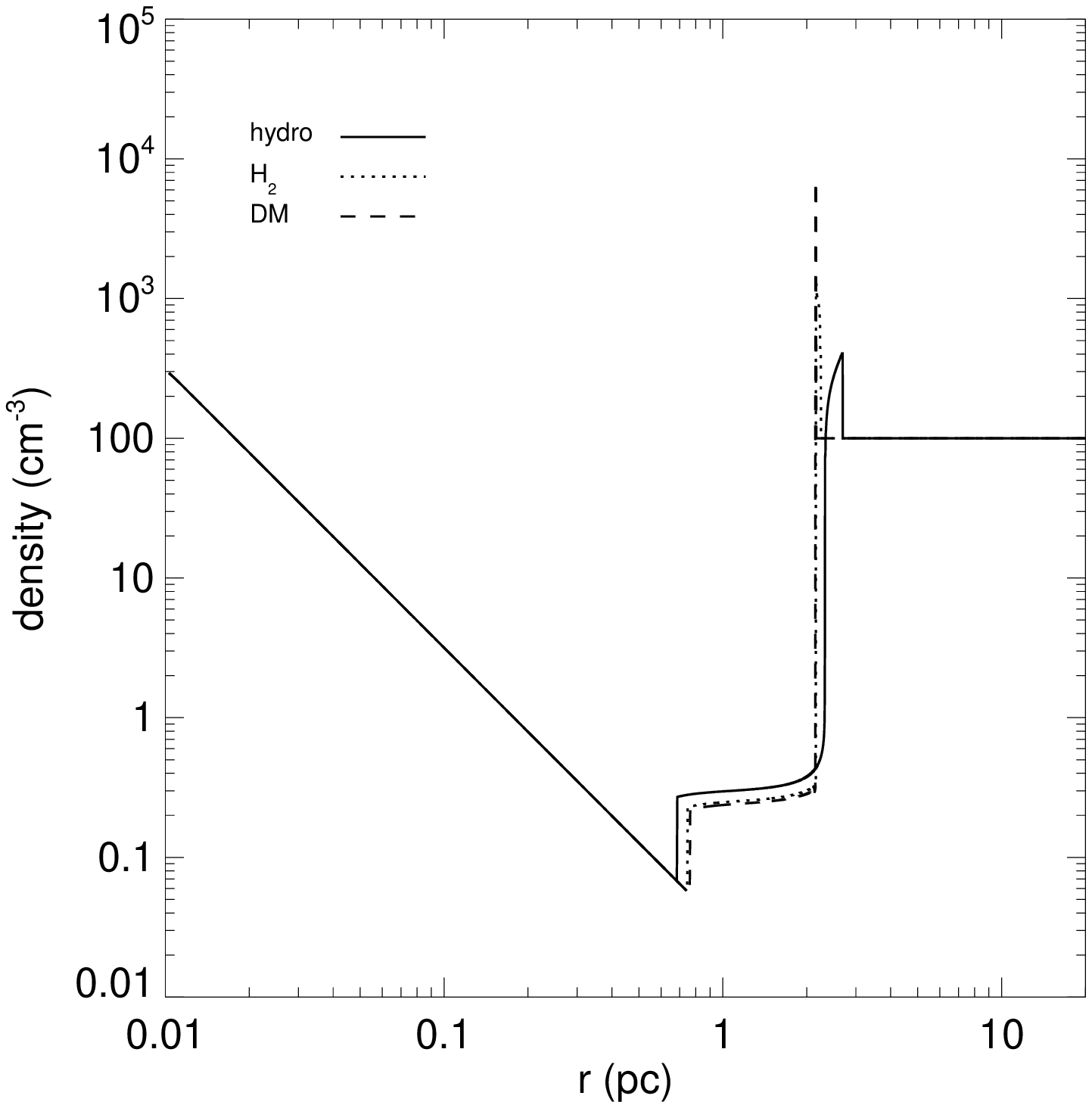}
\epsscale{2.0}
\caption{Left panel: wind-blown bubbles at 6.25 $\times$ 10$^4$ yr for $\dot{m}_w = 10^{-5}$ M$_{
\odot}$/yr and $v_w = 10^3$ km/s for four fiducial ambient densities, 10, 100, 100 and 1.8 $\times$ 
10$^4$ cm$^{-3}$.  These profiles include no radiative cooling. As can be seen in the plots, densities 
within $\sim$ 1 pc of the star depend only on $\dot{m}$ and $v_w$ for $n \lesssim$ 100 cm$^{-3}$.  
Right panel: structure of the shell at 1.25 $\times$ 10$^4$ yr in an ambient density of 100 cm$^{-3}$
with no cooling, H$_2$ cooling and fine structure cooling by C, O, N, Si and Fe at $Z =$ 0.1 $Z_{
\odot}$ (from the Dalgarno--McCray, or DM, cooling curves).  Radiative cooling flattens the shell 
plowed up by the wind into a cold dense structure, with no effect on the free-streaming region in the 
vicinity of the star.  Note also that efficient cooling in the shell also radiates away some of its thermal 
energy and slows its advance.} \vspace{0.1in}
\label{figure1_windBubbles}
\end{figure*}

The wind bubble has the classic two-shock structure first described by \citet{castorEA75} and \citet{weaverEA77}.  As we show in Fig.~\ref{figure1_windBubbles}, one shock forms at the interface between the emergent wind and the surrounding gas as it is swept up at speeds greatly exceeding the sound speed of the gas. As gas accumulates on the bubble, the shock detaches from the wind and moves ahead of it,  forming an intervening shell of dense postshock gas. At the same time, the expansion of the bubble evacuates a cavity into which the wind freely streams.  Since the shell moves more slowly than the wind, a termination shock also forms where the wind piles up against the inner surface of the shell.  If gas in the shell can radiatively cool, it flattens into a cold, dense structure that is prone to fragmentation into clumps.

We first performed four tests with steady winds of $\dot{m}_w$ = 10$^{-5}$ M$_{\odot}$/yr and $v_w$ = 1000 km/s in uniform densities $n= $ 10, 100, 1000 and 1.8 $\times$ 10$^4$ cm$^{-3}$ to investigate how local densities govern the radius of the bubble at intermediate times. For simplicity, these calculations were done with no chemistry or radiative cooling.  As we show in the left panel of Fig.~\ref{figure1_windBubbles}, ambient density governs only how far the bubble is driven from the star, not the profile of the free-streaming region in the immediate vicinity of the star, which is determined only by $\dot{m}$ and $v_{w}$ (and  fluctuations thereof). 

We find that radiative cooling has a dramatic effect on the structure of the primary shell but no influence on the flow up to the shell, as we show in the right panel of Fig.~\ref{figure1_windBubbles} for the same wind and $n =$ 100 cm$^{-3}$.  The three plots show the structure of the bubble with no cooling, H$_2$ cooling, and fine-structure cooling due to metals at $Z = $ 0.1 $Z_{\odot}$.  The cooling flattens the shell into a cold dense structure and radiates away some of its thermal energy, slightly retarding its advance.  The free-streaming zone is again unaffected because the wind velocity ensures that the termination shock is at least 0.6 pc from the star by the end of the simulation.  Thus, chemistry and cooling will clearly cause any subsidiary shells ejected by the progenitor at late times to be much thinner and denser, with potentially important consequences for the propagation of the jet.

\section{Circumburst Density Profiles of Collapsars and He Mergers}

\begin{figure}[t]
\centering
\begin{tabular}{c}
\epsfig{file=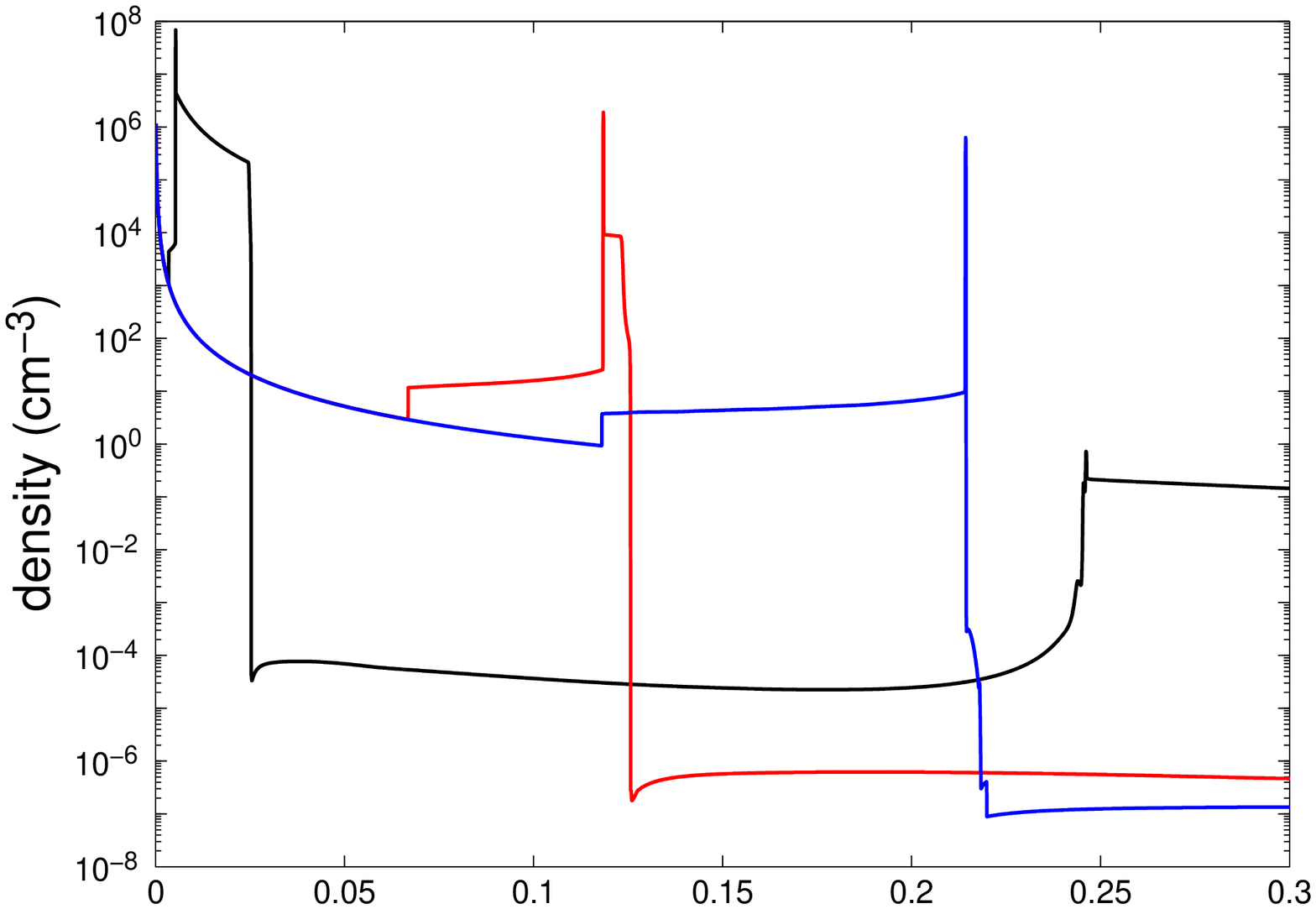, width=0.5\linewidth,clip=} \\
\epsfig{file=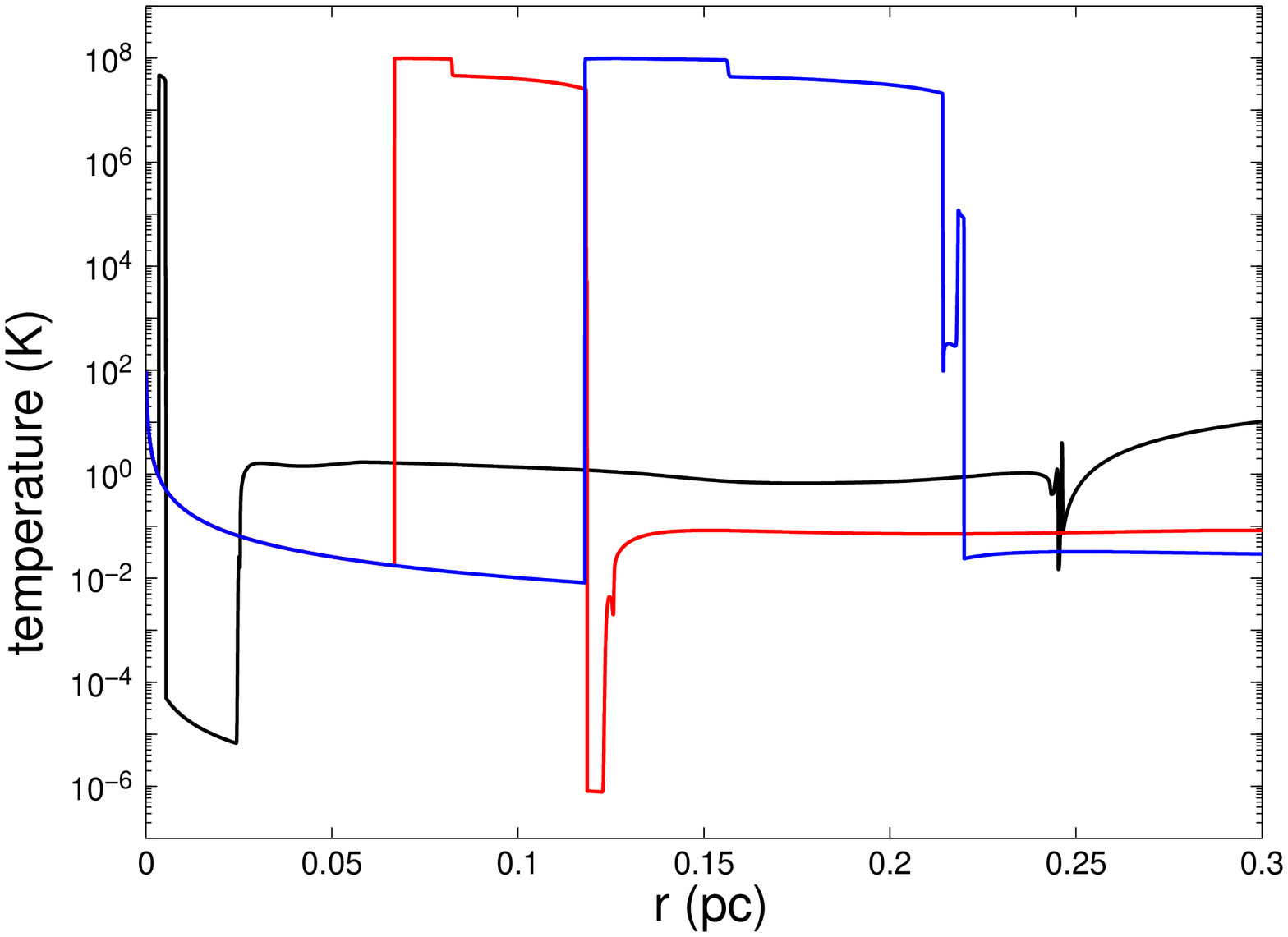, width=0.5\linewidth,clip=} \\
\end{tabular}
\caption{Density profiles (a) and temperature profiles (b) for a 100 yr outburst with $\dot{m}_b =$ 
10$^{-2}$ M$_{\odot}$/yr in a stellar wind with  $\dot{m}_w =$ 10$^{-5}$ M$_{\odot}$/yr.  Black:  
120 yr; red:  600 yr; blue:  1000 yr.     \label{figure2_exampleProfiles}}
\end{figure}

We now consider the more complicated structures in the immediate vicinity of the star at the time of the GRB.  We examine a fiducial case from our grid of shell models, the 100 yr outburst in a stellar wind with $\dot{m}_w = $ 10$^{-5}$ M$_{\odot}$/yr, whose density and temperature profiles we show at 120, 600 and 1000 yr in Fig.~\ref{figure2_exampleProfiles}.  As the shell emerges from the star, it promptly cools to extremely low temperatures, as we show in panel B of Fig.~\ref{figure2_exampleProfiles} at 120 yr.  This happens because fine-structure cooling timescales are less than a year in the dense 0.1 \ensuremath{Z_\odot}\ shell.  At the same time, the fast wind detaches from and races ahead of the shell, as we show in panel A of Fig.~\ref{figure2_exampleProfiles}. This creates a zone of rarefaction into which the shell freely streams, as shown by its roughly $r^{-2}$ density profile. 

As the wind pulls away from the shell, the abrupt adiabatic expansion causes temperatures at its inner edge to drop sharply.  The expansion of the rarefied region likewise causes its temperatures to fall to $\sim$ 1 K.  In reality, gas temperatures cannot fall below the cosmic microwave background temperature at this epoch, $T_{CMB} = 2.73 (1+z)$ K, but this is not included in our simulations.  We also ignore any heating of the gas due to other external processes such as cosmic rays or photons from the progenitor itself.  Although these processes could conceivably alter the density profile in the regions of free-streaming wind if those regions were to become ionized, the temperature of the shell is set by the conversion of the free-streaming wind's kinetic energy into thermal energy at the shell's inner edge (as well as through radiative cooling throughout its interior).  Any modifications of the shell structure due to heating by other processes is negligible. 

The density dip at the beginning of the rarefaction region and the density jump at the rear edge of the detached wind are complementary and due to mass conservation.  When the fast wind breaks away from the shell a thin shell of dense gas from its outer layers breaks off with it, leaving a thin layer whose density is even lower than that of the rarefied region and which remains a persistent feature of the flow out to $\sim$ 500 yr.  A transient structure with multiple strong density jumps results: the emerging massive shell and the inner edge of the rapidly receding wind separated by an intervening low density region.  The shell soon begins to plow up the low-density gas, as can be seen in the density bump just beyond its leading edge.  

After the shell has been fully ejected (and the fast wind has exited the grid), it soon evolves into a wind bubble as the fast wind piling up at its inner surface forms a termination shock that detaches and recedes from it in the frame of the shell.  The termination shock heats the gas piling up at the inner surface of the shell to $\sim$ 10$^8$ K while the shell itself remains cold due to fine-structure cooling.  Densities in the postshock gas between the termination shock and the shell become roughly uniform because sound speeds in the shell are $\sim$ 1100 km/s and it is nearly isothermal, so pressure gradients in this region arising from any initial density gradients across it are erased by acoustic waves on timescales that are short in comparison to the expansion times of the shell. Temperatures in the shell are lower than those of canonical wind bubbles because the ambient density is too low to heat the shell and activate H and He line cooling, in contrast to Fig.~\ref{figure1_windBubbles}. Strong bremsstrahlung x-ray flux from shocked gas at the inner surface of the shell likely ionizes it to some degree, but probably does not otherwise alter its properties because, at these photon energies, most of the kinetic energy of the photoelectrons goes into secondary ionizations rather than heat \citep[e.g.][]{shullvanSteenberg85,ricottiEA01,ricottiEA02,ricottiEA05}.  By 760 yr the shell has swept up enough low density gas to form a shock, which soon separates from the shell and advances beyond it as shown in the density and temperature plots at 1000 yr.  This feature is temporary: the shell eventually subsumes the secondary shock as it expands.  A free-streaming region forms behind the termination shock and extends to $\sim$ 0.1 pc by 1000 yr.  Part of the reason the shell drives a shock at intermediate times is that it accelerates as it is pushed by the fast wind and as it expands into low densities that continue to fall over time, as we show in all three density plots.  A velocity gradient develops across the shell as its inner and outer surfaces accelerate to 220 km/s and 280 km/s respectively over 1000 yr. 

Radiative cooling flattens the ejected shell as it expands into the surrounding medium.  Its width at first is 0.02 pc at 100 yr but later decreases essentially to the resolution limit of the grid by 1000 yr.  This is not unexpected:  when a dense shell driven by winds or swept up by a supernova can radiatively cool, its width usually drops to a few mesh zones, with the number of zones being determined by the numerical viscosity of the hydrodynamic algorithm.  All seven models exhibit the same general evolution, with the only variations being in the thickness of the shell over time, its peak densities, and its location on the grid at the time of the GRB.  Depending on when the explosion occurs, the jet will encounter three density jumps or four, with jump ratios of up to 10$^{10}$.  The thickness and average density of the shell varies from 0.02 to 0.0001 pc and 10$^{5}$ - 10$^{9}$ cm$^{-3}$ over 1000 yr.    

All our models of massive shell ejection follow the evolutionary sequence we have just described, with the only differences being in the magnitudes of density jumps, the mass and thickness of the shell at a given time, and the positions of the regions on the grid.


\section{Afterglow Light Curves}

We now discuss the general method we have devised for computing light curves for relativistic jets in the complicated wind structures we have modeled in ZEUS-MP.

\subsection{Blast Wave Hydrodynamics}

In the canonical fireball model, gamma ray bursts are modelled as initially highly-relativistic, adiabatic jets that propagate outward into an ambient medium.  We will assume that the jet expands adiabatically, i.e., only a very small fraction of the total burst energy is available to the electrons to be radiated away, and that all of the circumburst material in the jet's path is swept up.  In the following discussion, primed quantities refer to the reference frame that is comoving with the jet, unprimed quantities with no subscript refer to the reference frame in which the ISM is at rest, and quantities with the subscript $\earth$ refer to the reference frame of an Earthbound observer.  

Energy conservation requirements yield a formula for the evolution of the jet as it propagates through the external medium.  In the case where the jet expands adiabatically,

\begin{equation}
\frac{d\Gamma}{dm} = -\frac{\hat{\gamma}\left(\Gamma^2 - 1\right) - \left(\hat{\gamma} - 1\right)\Gamma\beta^2}{M_\text{ej} + m\left[2\hat{\gamma}\Gamma - \left(\hat{\gamma} - 1\right)\left(1 + \Gamma^{-2}\right)\right]},
\label{GammaEquation}
\end{equation}


\noindent where $\Gamma$ is the Lorentz factor of the jet, $M_\text{ej}$ is the initial mass of the jet ejecta, m is the total mass that has been swept up by the jet, $\beta = \left(1-\Gamma^{-2}\right)^{1/2}$ in the normalized bulk velocity, and $\hat{\gamma}$ is the adiabatic index \citep{peer12}.

The high resolution of our simulations ($10^{-4}\ \text{pc}$) allows us to take the density to be constant across each mesh zone. The total mass swept up by the jet by the time it reaches grid point $n$ is then approximately

\begin{equation}
M(r) = \frac{4}{3}\pi\left(\rho_1 r_1^3 + \sum_{i=2}^{n}{\rho_i\left(r_i^3 - r_{i-1}^3\right)} \right),
\end{equation}
where $r_i$ and $\rho_i$ are the radius and density of the $i$th grid point, respectively.  The time $t_\text{obs}$ at which a photon emitted at the shock boundary reaches an observer along the line of sight can be calculated by integrating equation 12 from \citet{huangEA99}:

\begin{equation}
t = \frac{1}{c}\int{\frac{dr}{\beta\Gamma\left(\Gamma + \sqrt{\Gamma^2 - 1}\right)}}\, \label{tObsEquation}
\end{equation}

\noindent where $\beta = v/c$ is the jet velocity and $c$ is the speed of light.

Many GRBs are thought to produce collimated jets rather than isotropic outflows.  The center of a relativistic jet of half-opening angle $\theta_j$ is not in causal contact with its edge until $t_\text{obs} = t_\text{jet}$, which is defined as the time where $\Gamma \simeq 1/\theta_j$.  The jet evolves in an identical manner as in the isotropic case until $t_\text{obs} = t_\text{jet}$, at which point the jet experiences rapid lateral expansion.  The increase in external mass being swept up by the jet after $t_\text{obs} = t_\text{jet}$ causes the jet to decelerate at an increasing rate.  Additionally, the afterglow's decreasing luminosity is no longer being partially offset by an increase in the size of the emitting region as seen by the observer, leading to a break in the light curve.  The overall evolution of the jet's angular size is 

\begin{equation}
\frac{d\theta_j}{dt} = \frac{c'_s\left(\Gamma + \sqrt{\Gamma^2 - 1}\right)}{r},
\end{equation}

where $c'_s$ is the comoving sound speed and $r$ is the radius of the jet \citep{huangEA00}.

\subsection{The Injection Break}

If we assume that a constant fraction $\epsilon_B$ of the total fireball energy is stored in magnetic fields, then the equipartition magnetic field strength at the shock boundary is \citep[i.~e.][]{panaitescuKumar00}:

\begin{equation}
\frac{B'^2}{8\pi} = 4\epsilon_Bm_pc^2n(r)(\Gamma-1)\left(\Gamma + \frac{3}{4}\right),
\end{equation}   

\noindent where $m_p$ is the proton mass and $n(r)$ is the number density of the medium at radius $r$.

The electrons that are injected into the shock are assumed to have a velocity distribution $N(\gamma) \propto \gamma^{-p}$ with a minimum Lorentz factor $\gamma_m$.  Electrons with a Lorentz factor $\gamma_e$ emit synchrotron radiation at a characteristic frequency \citep{rybickiLightman79}:

\begin{equation}
\nu(\gamma_e) = \Gamma\gamma_e^2\frac{q_eB}{2\pi m_ec}. \label{characteristicFrequencyEquation}
\end{equation}
  
\noindent The injection break, $\nu_m$, corresponds to the characteristic frequency at which the electrons having the minimum Lorentz factor radiate.  The minimum Lorentz factor is \citep{sariEA98}

\begin{equation}
\gamma_m = \left(\frac{p - 2}{p - 1}\right)\frac{m_p}{m_e}\epsilon_e\left(\Gamma - 1\right), \label{minLorentzFactorEquation}
\end{equation}

\noindent where $m_e$ is the electron masses, respectively.  Note that equation \ref{minLorentzFactorEquation} is only valid in the relativistic limit.  Substituting equation \ref{minLorentzFactorEquation} into equation \ref{characteristicFrequencyEquation} will therefore only yield $\nu_m$ as long as $\Gamma \gtrsim 2.0$.  \citet{frailEA00} show that, for $\Gamma \lesssim 2.0$, the injection break frequency follows the simple relation $\nu_m \propto t^{-3}$.

\subsection{The Cooling Break}

Relativistic electrons in the shock cool radiatively through inverse Compton (IC) scattering and synchrotron emission on a co-moving frame timescale

\begin{equation}
t'_\text{rad}(\gamma) = \frac{6\pi}{Y+1}\frac{m_ec}{\sigma_e\gamma B'^2}, \label{remnantAgeEquation}
\end{equation}

\noindent where $Y$ is the Compton parameter and $\sigma_e$ is the Thompson scattering cross section \citep{panaitescuKumar00}.  An electron with Lorentz factor $\gamma_c$ cools radiatively on a timescale equal to the current age of the remnant.  Solving Eqn. \ref{remnantAgeEquation} for $\gamma_c$, we find that the Lorentz factor for electrons that cool on a timescale equal to the observer-frame age of the remnant is

\begin{equation}
\gamma_c = \frac{6\pi m_ec^2}{B'^2\sigma_e(Y + 1)t'}. \label{coolingBreakEquation}
\end{equation}

\subsubsection{Fast-Cooling Electrons}

Electrons in the GRB jet can cool by adiabatic expansion of the gas or by emission of radiation.  When the cooling timescale for electrons with Lorentz factor $\gamma_m$ is less than the age of the jet ($\nu_c < \nu_m$, where $\nu_c$ is the frequency of the cooling break) the electrons in the jet lose a significant portion of their energy through emission of radiation and are said to be radiative, or fast-cooling.  Conversely, if the cooling timescale is 
greater than the age of the jet ($\nu_c > \nu_m$) the electrons do not lose significant energy to radiation and are said to be adiabatic, or slow-cooling.

To calculate the Compton parameter, $Y$, we only account for one upscattering of the synchrotron photons.  If the injected electrons are fast-cooling and the frequency of the absorption break $\nu_a < \text{min}(\nu_m, \nu_c)$, then $Y$ can be approximated by \citet{panaitescuMeszaros00}:

\begin{equation}
Y_r = \gamma_\text{m}\gamma_\text{c}\tau_\text{e}, \label{radiativeComptonParameterEquation1}
\end{equation}

\noindent where a constant of order unity has been ignored and $\tau_e$ is the optical depth to electron scattering, given by

\begin{equation}
\tau'_e = \frac{\sigma_eM(r)}{4\pi m_\text{p}r'^2}.
\end{equation}

\noindent The medium becomes optically thick to synchrotron self-absorption at the absorption break frequency $\nu_a$.  When both the injection break and the cooling break lie in the optically thick regime, $Y$ becomes

\begin{equation}
Y_r = Y_* = \tau'_e\left(C_2^{2-p}\gamma_c^7\gamma_m^{7(p-1)}\right)^{1/(p+5)},
\end{equation}

\noindent where $C_2 \equiv 5q_e\tau'_e/\sigma_eB'$ \citep{panaitescuMeszaros00}. 

\subsubsection{Slow-Cooling Electrons}

If the electrons are slow-cooling, then $Y$ becomes

\begin{equation}
Y_a = \tau'_\text{e}\gamma_\text{i}^{p-1}\gamma_\text{c}^{3-p},
\end{equation}

\noindent as long as $\nu_a < \text{min}(\nu_m, \nu_c)$ \citep{panaitescuMeszaros00} and $1 < p < 3$.  Once again we have ignored a constant of order unity.  If both the injection and cooling breaks lie in the part of the spectrum that is optically thick to synchrotron self-absorption, then $Y$ is identical to the corresponding fast-cooling case and

\begin{equation}
Y_a = Y_*.
\end{equation}

\subsection{The Absorption Break}

At lower frequencies, the medium through which the jet propagates becomes optically thick to synchrotron self-absorption.  The result is a transition to a $F_\nu \propto \nu^2$ drop-off in the flux at some absorption break frequency $\nu_a$ where the optical depth to self-absorption is $\tau_\text{ab} = 1$.  The frequency of the absorption break depends on the electron cooling regime (fast or slow) and on the order and values of both the injection and cooling breaks.  In the fast-cooling regime, \citet{panaitescuMeszaros00} find that

\begin{equation}
\nu'_\text{a, fast-cooling} = 
\begin{cases}
C_2^{3/10}\gamma_c^{-1/2}       ,                 & \gamma_a < \gamma_c < \gamma_m \\
\left(C_2\gamma_c\right)^{1/6} ,                  & \gamma_c < \gamma_a < \gamma_m \\
\left(C_2\gamma_c\gamma_m^{p-1}\right)^{1/(p+5)}, & \gamma_c < \gamma_m < \gamma_a, \\
\end{cases}
\end{equation}

\noindent whereas in the case where the electrons are slow-cooling

\begin{equation}
\nu'_\text{a, slow-cooling} = 
\begin{cases}
C_2^{3/10}\gamma_m^{-1/2}       ,                  & \gamma_a < \gamma_m < \gamma_c \\
\left(C_2\gamma_m^{p-1}\right)^{1/(p-4)} ,        & \gamma_m < \gamma_a < \gamma_c \\
\left(C_2\gamma_m^{p-1}\gamma_c\right)^{1/(p+5)}, & \gamma_m < \gamma_c < \gamma_a. \\
\end{cases}
\end{equation}

\subsection{Light Curves}

In order to produce light curves, we must first find the time dependence of $\Gamma(r)$, $n(r)$, and $M(r)$.  Equation \ref{GammaEquation} can be solved numerically for $\Gamma(r)$, and equation \ref{tObsEquation} can then be used to relate the observer time $t_\text{obs}$ to the jet position $r$, allowing us to rewrite the equations defining the three break frequencies in terms of $t_\text{obs}$, $\Gamma(t_\text{obs})$, $n(t_\text{obs})$, and $M(t_\text{obs})$.  Given the three break frequencies and the peak flux density, analytical light curves can then be calculated that are valid from the radio to the $\gamma$-ray regions of the spectrum.  If $\nu_a < \text{min}(\nu_m, \nu_c)$, then the peak flux density $F_{\nu,\earth}^\text{max}$ occurs at the injection break if $\nu_\text{m} < \nu_\text{c}$ and at the cooling break if $\nu_\text{m} > \nu_\text{c}$:

\begin{equation}
F_{\nu,\earth}^\text{max} = \frac{\sqrt{3}\phi_\text{p}}{4\pi D^2}\frac{e^3}{m_\text{e}c^2}\frac{\Gamma B'M(r)}{m_\text{p}},
\end{equation}

\noindent where $\phi_\text{p}$ is a factor calculated by \citet{wijersGalama99} that depends on the value of p, and $D = (1+z)^{-1/2}D_l$, where $D_l$ is the luminosity distance to the source \citep{panaitescuKumar00}.  The flux at any frequency $\nu$ (ignoring relativistic beaming and the spherical nature of the emitting region) has been derived by \citet{sariEA98} and \citet{panaitescuKumar00}.  

\subsubsection{Fast-Cooling Electrons}

When the electrons are in the fast-cooling regime, the peak flux density occurs at the cooling break as long as $\nu_a < \nu_c$:

\begin{equation}
F_{\nu, \earth} = F_{\nu,\earth}^\text{max}
\begin{cases}
(\nu/\nu_a)^2(\nu_a/\nu_c)^{1/3},       &         \nu < \nu_a \\
(\nu/\nu_c)^{1/3},                      & \nu_a < \nu < \nu_c \\
(\nu/\nu_c)^{-1/2},                     & \nu_c < \nu < \nu_m \\
(\nu/\nu_m)^{-p/2}(\nu_m/\nu_c)^{-1/2}, & \nu_m < \nu.
\end{cases}
\label{fluxEquation1}
\end{equation}

\noindent If the medium is optically thick to synchrotron self-absorption at the cooling break frequency, then the maximum flux moves to the absorption break frequency.  Between the absorption break and the cooling break, $F_\nu \propto \nu^{5/2}$ but it becomes $\propto \nu^{2}$ below the cooling break:

\begin{equation}
F_{\nu, \earth} = F_{\nu,\earth}^\text{max}
\begin{cases}
(\nu/\nu_c)^2(\nu_c/\nu_a)^{5/2},       &         \nu < \nu_c \\
(\nu/\nu_a)^{5/2},                      & \nu_c < \nu < \nu_a \\
(\nu/\nu_a)^{-1/2},                     & \nu_a < \nu < \nu_m \\
(\nu/\nu_m)^{-p/2}(\nu_m/\nu_a)^{-1/2}, & \nu_m < \nu.
\end{cases}
\end{equation}

In the canonical afterglow models that assume a uniform density environment, the cooling break and the injection break move to lower frequencies with time.  Eventually, both the cooling break and the injection break can lie below the absorption break, but far too late in the evolution of the burst to be relevant to anything but the radio afterglow, and long after 
the time at which the electrons in the jet have transitioned to the slow-cooling regime.  In our more realistic density profile models, the extremely high density encountered by the jet as it passes through the thick shell causes it to abruptly transition from highly relativistic to Newtonian expansion.  The decrease in $\Gamma$ leads to a sharp drop in the 
injection break frequency, while the increased medium density leads to a larger magnetic field strength, which in turn causes a drop in the cooling break frequency.  The result is that the absorption break frequency can be several orders of magnitude higher than the cooling and injection break frequencies as the jet traverses the thick shell.  Multiple transitions 
between fast and slow electron cooling can also occur.  In the vicinity of the thick shell, when $\nu_a > \nu_m$ and the electrons are in the fast-cooling regime:

\begin{equation}
F_{\nu, \earth} = F_{\nu,\earth}^\text{max}
\begin{cases}
(\nu/\nu_c)^2(\nu_c/\nu_a)^{5/2}, &         \nu < \nu_c \\
(\nu/\nu_a)^{5/2},                & \nu_c < \nu < \nu_a \\
(\nu/\nu_a)^{-p/2},								& \nu_m < \nu.
\end{cases}
\label{fluxEquation3}
\end{equation}

\subsubsection{Slow-Cooling Electrons}

Our models yield the same flux as the canonical wind models until the jet encounters the shocked wind that has piled up behind the thick ejecta shell.  If it does not encounter the shocked wind in the first few hours, the electrons in the shock transition to the slow-cooling regime, with $\nu_a \ll \nu_m$ and

\begin{equation}
F_{\nu, \earth} = F_{\nu,\earth}^\text{max}
\begin{cases}
(\nu/\nu_a)^2(\nu_a/\nu_m)^{1/3},           &         \nu < \nu_a \\
(\nu/\nu_m)^{1/3},                          & \nu_a < \nu < \nu_m \\
(\nu/\nu_m)^{-(p-1)/2},                     & \nu_m < \nu < \nu_c \\
(\nu/\nu_c)^{-p/2}(\nu_c/\nu_m)^{-(p-1)/2}, & \nu_c < \nu.
\end{cases}
\end{equation}

\noindent In the case where $\nu_m < \nu_a < \nu_c$,

\begin{equation}
F_{\nu, \earth} = F_{\nu,\earth}^\text{max}
\begin{cases}
(\nu/\nu_m)^2(\nu_m/\nu_a)^{5/2},           &         \nu < \nu_a \\
(\nu/\nu_a)^{5/2},                          & \nu_a < \nu < \nu_m \\
(\nu/\nu_a)^{-(p-1)/2},                     & \nu_m < \nu < \nu_c \\
(\nu/\nu_c)^{-p/2}(\nu_c/\nu_a)^{-(p-1)/2}, & \nu_c < \nu.
\end{cases}
\end{equation}

As noted earlier, as the jet passes through the thick shell, it can experience multiple transitions between fast and slow electron cooling.  When the electrons are in the slow-cooling regime and $\nu_a > \nu_c$,

\begin{equation}
F_{\nu, \earth} = F_{\nu,\earth}^\text{max}
\begin{cases}
(\nu/\nu_m)^2(\nu_m/\nu_a)^{5/2}, &         \nu < \nu_m \\
(\nu/\nu_a)^{5/2},                & \nu_m < \nu < \nu_a \\
(\nu/\nu_a)^{-p/2}, 							& \nu_a < \nu.
\end{cases}
\label{fluxEquation6}
\end{equation}

\subsection{Spherical Emission and Beaming}

The spherical nature of GRB afterglow emission can substantially alter the observed light curve \citep{fenimoreEA96}.  The burst ejecta is initially ultra-relativistic, with $100 \lesssim \Gamma \lesssim 1000$, meaning that radiation that is emitted along the observer's line of sight will be beamed toward the observer more than radiation that is emitted away from the line of sight.  There will also be a delay in the arrival time of photons that are emitted from regions of the jet that lie away from the observer's line of sight, since these regions are further away from the observer \citep{fenimoreEA96}.  The overall effect will be to increase the total observed flux at early times, and to make the light curve broader and more smooth.  






\begin{figure}[t]
\centering
\begin{tabular}{c}
\epsfig{file=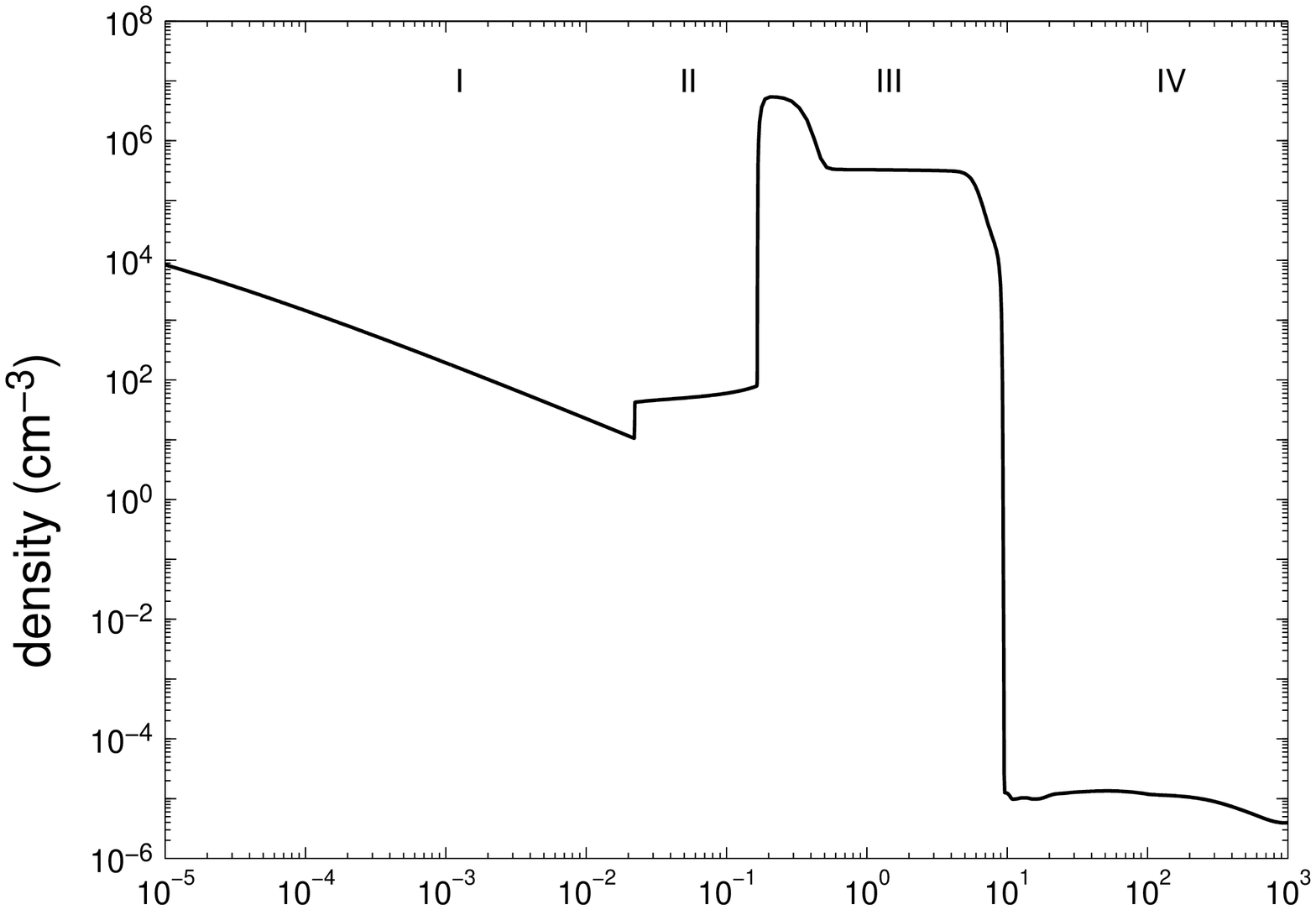, width=0.5\linewidth,clip=} \\
\epsfig{file=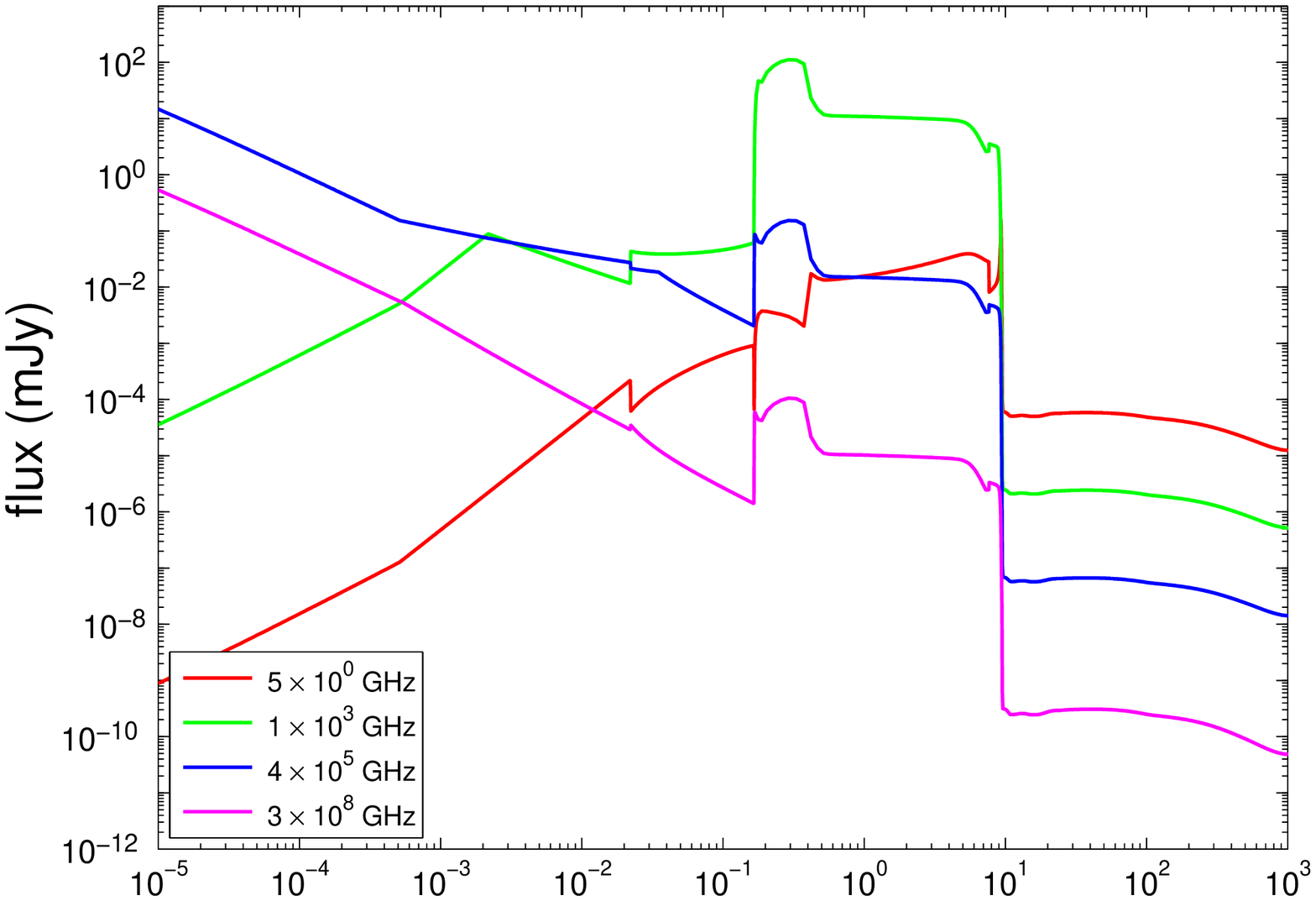, width=0.5\linewidth,clip=} \\
\epsfig{file=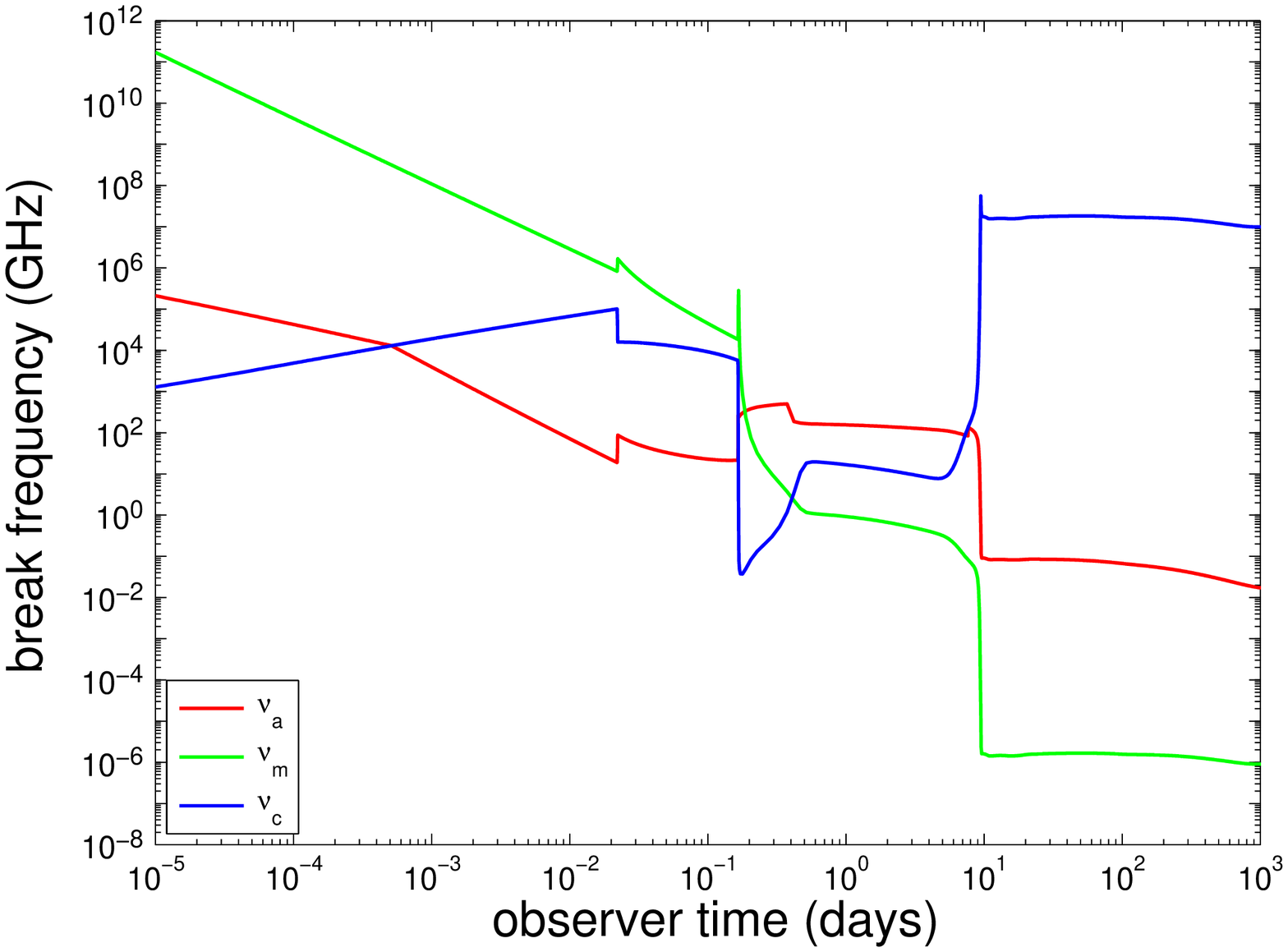, width=0.5\linewidth,clip=} \\
\end{tabular}
\caption{Light curves and break frequencies for a GRB in a dense shell.  Panel (a): densities encountered by the jet over time; panel (b):  synchrotron light curves; panel (c): break frequencies.  Region V, a second region of unshocked wind, is not shown here because the fast detached wind has exited the grid.    \label{figure3_lightCurves}}
\end{figure}

\subsection{Imprint of Dense Shells on GRB light curves}

The shell and wind enclosing the GRB can be partitioned into four distinct regions, which we show in Fig.~\ref{figure3_lightCurves}a for a progenitor with a 0.1 M$_{\odot}$ shell ejected 100 yr before the burst and mass loss rates of 10$^{-6}$ M$_{\odot}$/yr before and after the ejection of the shell.  A fifth region, not shown in Fig.\ \ref{figure3_lightCurves}, also exists, but is located at such large radii at the time of the GRB that the jet does not encounter it until very late times.  Even so, this region is discussed below for completeness.  We now examine the imprint of each region on the GRB light curves and break frequencies, which we show in Figures ~\ref{figure3_lightCurves}b and \ref{figure3_lightCurves}c.

\subsubsection{Region I -- Unshocked Wind}

Region I is the unshocked wind blown by the progenitor after the ejection of the shell and prior to the burst.  Not surprisingly, the light curve in this region is what would be expected for a $\rho(r) \propto r^{-2}$ wind profile (Fig.~\ref{figure3_lightCurves}b).  The jet electrons are initially in the fast cooling regime, and the absorption break frequency is generally higher than the cooling break frequency for the first few seconds, as is evident in Fig.~\ref{figure3_lightCurves}c.  

\subsubsection{Region II -- Shocked Wind}

Region II is the shocked wind that has piled up behind the shell, where the density jumps by about an order of magnitude and transitions from an $r^{-2}$ density profile to a nearly flat one (Fig.~\ref{figure3_lightCurves}a).  Although the increased density has little effect on the injection break frequency, both the cooling break and the absorption break each change abruptly by approximately an order of magnitude (Fig.~\ref{figure3_lightCurves}c).  There is a moderate change in the flux at all frequencies, though the magnitude of the change at each frequency and whether it is an increase or a decrease depends upon the values of the break frequencies.  Because region II's imprint on the light curve is modest at best even when the emitting region's spherical nature is ignored, it is likely that the inclusion of spherical emission would eliminate the imprint entirely.  This is because the jet is still highly relativistic when it enters region II, and the observer does not see the radiation that is emitted when the jet enters region II all at once.  The result is that the small bump in the light curve that occurs at the interface between regions I and II would be smoothed out, and region II would have essentially no effect on the light curve.

Depending on the progenitor's mass-loss rate, the mass of the shell, and the delay between the expulsion of the shell and the burst, the jet can arrive at this region in several minutes in the limit of low mass-loss rates, high shell masses and short delays between the shell ejection and the burst to a year in the limit of high loss rates, low shell mass and long delays between the shell ejection and the burst.  In Fig.~\ref{figure4_ShellB004} we show how the time between shell ejection and the burst governs when the jet reaches region II, and hence when the light curves and breaks would diverge from those expected for $r^{-2}$ wind profiles or uniform density fields. In Fig.~\ref{figure5_modelComparison}a we show how mass-loss rates and shell masses impact these arrival times.

\subsubsection{Region III -- Dense Shell}

\begin{figure}[h]
\centering
\includegraphics[width=0.5\textwidth]{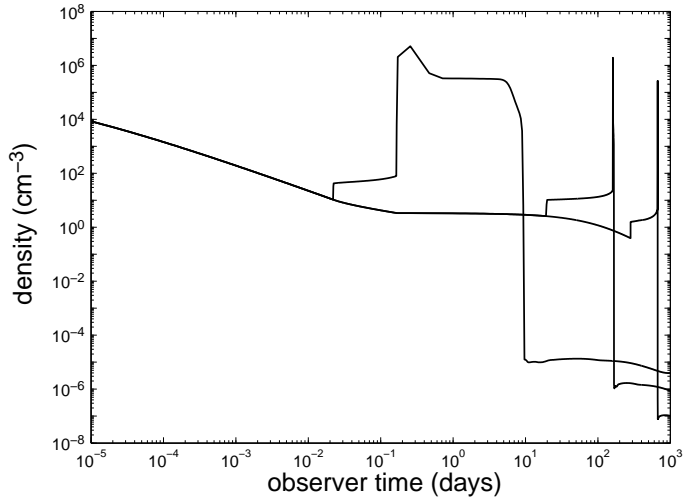}
\caption{Density encountered by the jet as it passes through the shell as a function of observer time $t_{obs}$ for shells ejected 100, 200, and 500 yr prior to the GRB (left to right).  Here, the progenitor had a mass loss rate of $10^{-6}$ M$_{\odot}$ yr$^{-1}$ and a hydrogen shell of mass 0.1 M$_{\odot}$. These plots, and our calculations, implicitly include the slowing of the jet in the dense shell.  \label{figure4_ShellB004}}
\end{figure}

Region III is the dense shell ejected by the progenitor.  The jet collides with it in about an hour to several years after the burst, depending upon the progenitor's wind mass-loss rate, the mass of the shell, and the delay between shell ejection and the burst (Figs.~\ref{figure4_ShellB004} \& \ref{figure5_modelComparison}a).  When the jet crosses into region III it abruptly becomes non-relativistic because of the large density jump there, which can be up to ten orders in magnitude (Fig.~\ref{figure5_modelComparison}a).  A bright, highly relativistic reverse shock almost certainly forms, which we have neglected for simplicity. 

The transit time through the shell depends on its mass and the time between its ejection and the burst.  Massive shells decelerate the jet more than less massive ones, increasing the time that the jet is within the shell. The time between the ejection and burst governs the degree to which radiative cooling flattens the shell into a thin, cold, dense structure, shortening the time the jet is inside the shell.  Both factors cause the crossing time to vary from as little as 
a day to as much as several hundred days (Fig.~\ref{figure5_modelComparison}a).  

The shell leaves a clear imprint on the spectrum of the jet.  Upon collision, there is a sharp drop in the cooling and injection break frequencies and an increase in the absorption break frequency.  This causes an abrupt increase in the flux at all frequencies (Fig.~\ref{figure5_modelComparison}b).  Inside the shell, the cooling break frequency rises and the magnetic field strength falls as the jet decelerates, and the injection break evolves to lower frequencies.  

The delay in arrival time of photons emitted away from the observer's line of sight will certainly decrease the imprint of the circumstellar shell on the light curve.  At x-ray frequencies and higher, where the imprint of the shell is least dramatic, there may be little if any increase in the flux at the time that the jet enters the shell, though there should be a break in the light curve at the time that the jet enters the shell where the flux will begin to decay at a slower rate.  The jet becomes nonrelativistic soon after it enters the circumstellar shell, greatly diminishing the importance of the spherical nature of the emitting region.  As the jet traverses the shell, the observed flux must therefore approach the value that it would have if spherical emission was ignored.  The maximum flux $F_\nu$ occurs at some frequency $\nu_{\rm{shell}}^{\rm{max}} = \rm{min}(\nu_m, \nu_c)$ in the circumstellar shell if $\nu_a < \rm{min}(\nu_m, \nu_c)$, and at $\nu_{\rm{shell}}^{\rm{max}} = \nu_a$ otherwise.  The increase in flux is largest for frequencies below $\nu_{\rm{shell}}^{\rm{max}}$, (Fig.~\ref{figure5_modelComparison}), and the shell's imprint on the lightcurve is also consequently the largest.  Even including the effect of spherical emission, a significant rebrightening should still occur for frequencies below $\nu_{\rm{shell}}^{\rm{max}}$ (typically optical or infrared frequencies and below).  While spherical emission will have a strong effect on the lightcurve at the time that the jet enters the shell, it will not significantly alter the light curve when the jet exits the shell because the jet is non-relativistic.  The abrupt drop in the flux that occurs when the jet exits the shell should therefore remain clearly visible in the light curve at all frequencies.

\subsubsection{Region IV -- Low-Density Cavity}

Region IV is the extremely low-density cavity created when the fast wind beyond the shell detaches from and races ahead of it.  The densities in this region are so low ($10^{-5} - 10^{-8}\ \text{cm}^{-3}$, Fig.~\ref{figure5_modelComparison}a) that the flux density drops at all wavelengths when the jet exits the shell (Figs.~\ref{figure3_lightCurves}b \& 
\ref{figure5_modelComparison}b).  By this stage, the jet has become non-relativistic, and photons emitted away from the line of sight are no longer significantly delayed with respect to those emitted by the portion of the jet that moves directly toward the observer.  The observer sees the entire leading edge of the jet reach region IV at nearly the same time, resulting in an abrupt drop in the flux of several orders of magnitude at all but radio wavelengths, where the drop is not quite as large.  Because the jet does not sweep up much material in this region, its magnetic field strength and velocity taper off slowly. Consequently, the break frequencies are roughly constant (Fig.~\ref{figure3_lightCurves}b) with $\nu_m \ll \nu_a \ll \nu_c$, and the spectrum is fairly constant for tens to hundreds of days (Figs.~\ref{figure3_lightCurves}b \& \ref{figure5_modelComparison}b).  The absorption break generally lies below 1 GHz so the flux density becomes negligible for frequencies above the radio band (Fig.~\ref{figure3_lightCurves}b).

\subsubsection{Region V -- Detached Wind}

Eventually, the jet crosses the rarefied region and catches up to the wind that preceded the ejection of the shell.  The detached wind, region V, exhibits a sudden density jump of several orders of magnitude followed by an $r^{-2}$ dropoff thereafter (black plot of Fig.~\ref{figure3_lightCurves}a).  Consequently, another reverse shock may form at the interface between regions IV and V. Unlike the reverse shocks expected between regions I and II and between regions II and III, this one is Newtonian.  We evolved the wind and the shell out to a radius of 0.3 pc, and only when the burst occurs within 100 years of shell ejection does the jet overtake the detached wind before it exits the grid.  Our models predict that the jet will not reach region V for at least several years after the burst, which is why we do not show it in Fig. \ref{figure3_lightCurves}a.

\section{Discussion and Conclusions}

Our calculations show that GRB light curves in dense shells exhibit clear departures from those in canonical winds and uniform densities.  These features can broadly discriminate between classes of GRB collapse scenarios.  Analytical models predict that light curves for bursts in uniform or $r^{-2}$ density profiles are piecewise power law segments separated by breaks.  In contrast, light curves for GRBs in dense shells ejected by the progenitor (predicted for both single and binary He mergers) initially follow those for simpler environments, but deviate from them in most cases on timescales of a few hours to a few days.  The first departure is a sudden change in the flux of about an order of magnitude when the jet enters the shocked wind piled up behind the dense shell. A second, much more significant departure occurs soon thereafter when the jet collides with the dense shell.   

The features discussed above are distinct signatures of a GRB in the dense shell associated with some collapse scenarios.  The GRB 091127 lightcurve may provide evidence for a jet encountering a thick shell of material.  \citet{filgasEA11} determined that the temporal evolution of the GRB 091127 cooling break frequency could only be consistent with the standard fireball model if the GRB occured in an $r^{11}$ medium.  Such a sharp increase in density with radius is easily produced in a wind bubble environment at the trailing edge of the shell.

Our work also provides an alternative mechanism for the bright flares that are sometimes seen in GRB lightcurves within the first few thousand seconds of the onset of the afterglow emission.  GRB 081029 produced a flare in the optical band that began at $\sim0.035$ days, peaked at about 0.07 days after a total increase in brightness of 1.1 magnitudes, and then slowly dimmed until $\sim0.2$ days \citep{nardiniEA11}.  The structure and duration of the GRB 081029 optical flare is somewhat similar to the feature in Fig. \ref{figure3_lightCurves}b that peaks at 0.3 days.  A less-intense flare was observed in the GRB 071112C X-ray afterglow.  The fact that no corresponding flare was detected in the optical was used by \citet{huangEA12a} to argue that different mechanisms were responsible for the time evolution of the X-ray and optical afterglows.  The simple power law structure of the optical light curve was shown to be consistent with the external shock model, whereas late injection and internal shocks inside the jet itself were invoked to explain the excess X-ray emission.  Our work shows that there can be prominent, chromatic features in the light curves due solely to variations in the circumburst medium.  For example, Fig.~3b clearly illustrates that the structure of the light curves vary with energy band, and that it is not necessary to invoke delayed energy injection or other effects to account for these differences.  A similar case is that of the ``giant'' X-ray flare of GRB 050502B, wherein the photon count rate was seen to increase by a factor of 500 at $345\pm30$s \citep{falconeEA06}.  Like GRB 071112C, GRB 050502B was not observed to have an optical counterpart \citep{falconeEA06}, which has been used as evidence of a late-injection event.  Our models can produce flares of comparable magnitude and duration, but can also explain the broad plateau in the X-ray flux that lasts from the end of the flare up until about a day after the burst.  A GRB jet that encounters a thick shell of circumburst material will produce a flare that will rapidly dim as the jet becomes nonrelativistic.  The flux then remains nearly constant until the jet emerges from the leading edge of the roughly uniform-density shell and encounters a sharp drop in the medium density. 

\begin{figure}[h]
\centering
\begin{tabular}{c}
\epsfig{file=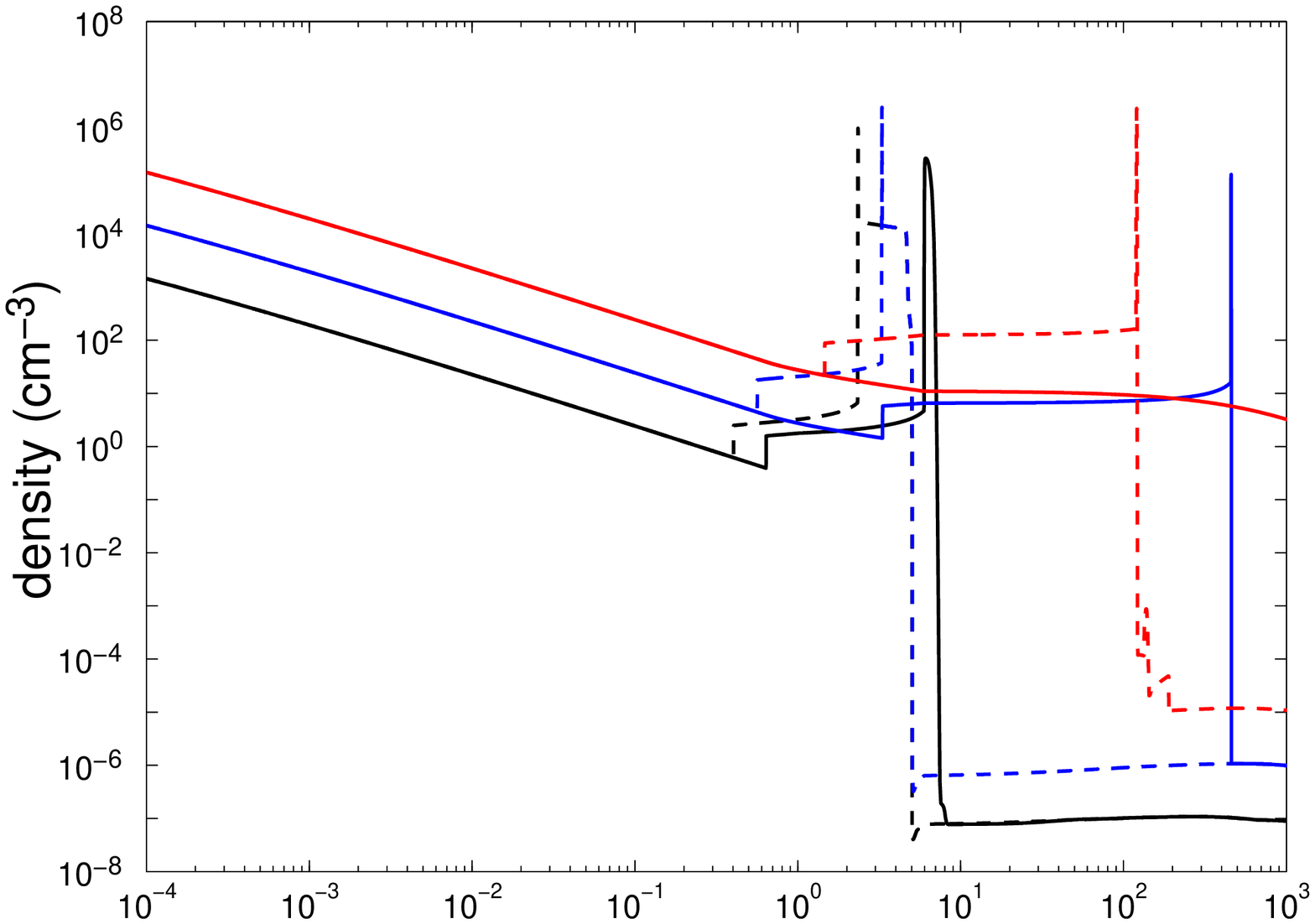, width=0.5\linewidth,clip=} \\
\epsfig{file=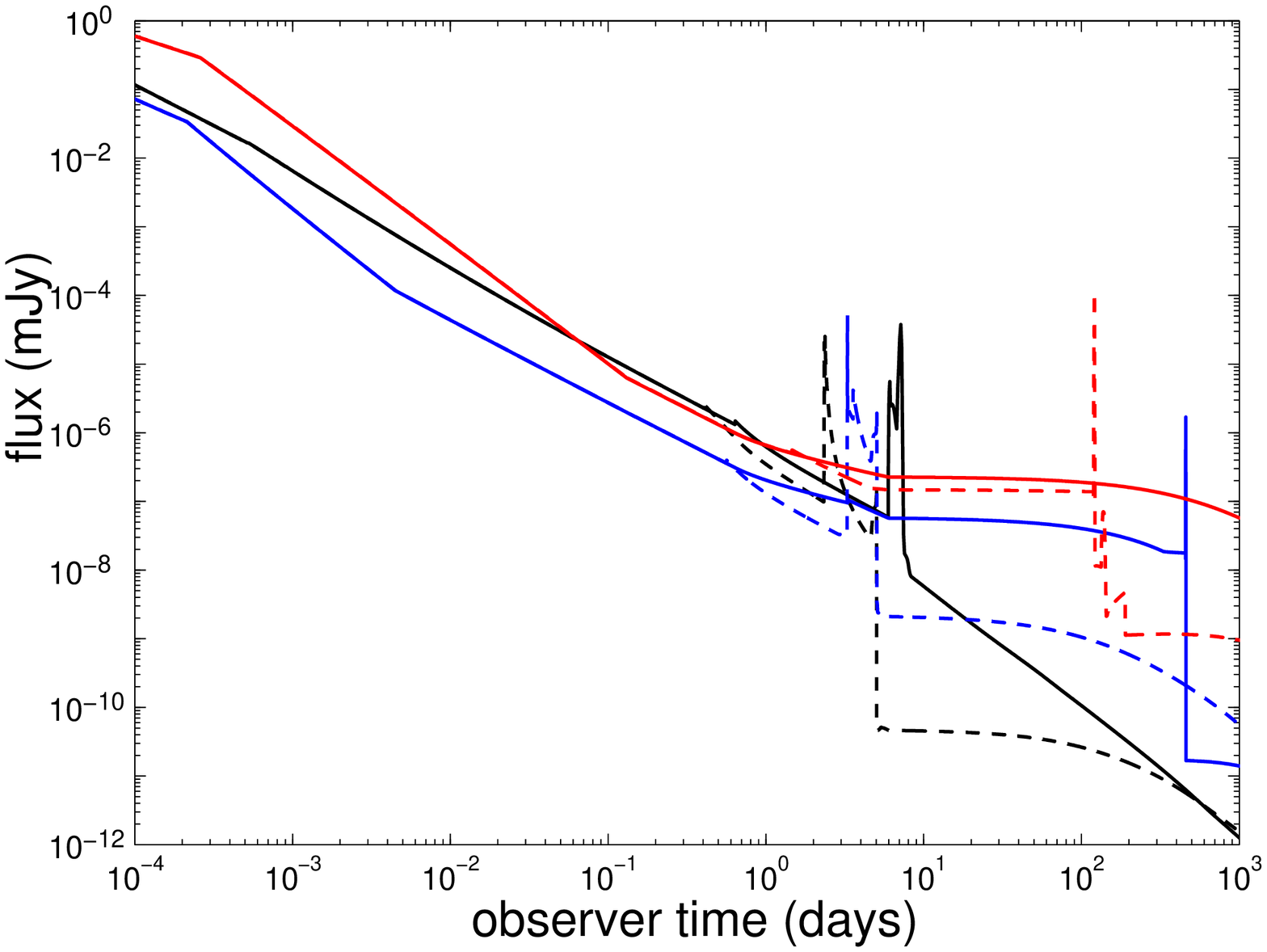, width=0.5\linewidth,clip=} \\
\end{tabular}
\caption{Density profiles (a) and gamma-ray light curves (b) ($3\times10^8$ GHz, or 1.24 keV) for a GRB occurring 500 yr after ejection of the progenitor's hydrogen shell.  Black:  ${\dot{m}}_w = 10^{-6}$ M$_{\odot}$ yr$^{-1}$; blue:  ${\dot{m}}_w = 10^{-5}$ M$_{\odot}$ yr$^{-1}$; red: ${\dot{m}}_w = 10^{-4}$ M$_{\odot}$ yr$^{-1}$.  Solid lines:  0.1 M$_{\odot}$ shells; dotted lines: 1 M$_{\odot}$ shells.    \label{figure5_modelComparison}}
\end{figure}

The unique imprint of dense shells on GRB light curves can be used to constrain their properties.  As shown in Fig.~\ref{figure5_modelComparison}b, the afterglow flux is somewhat sensitive to the ambient density out to $\sim$ a day for a shell ejected 500 yr before the GRB.  The light curve is also sensitive to the mass of the shell.  The crossing time in a 0.1 M$_\odot$ shell is a day or less, after which there is a sharp drop in the flux.  On the other hand, a 1.0 M$_\odot$ shell has crossing times of tens to hundreds of days, which creates a broad plateau in the light curve.  The mass-loss rate of the progenitor and the time at which the jet reaches the shell are also manifest in the gamma-ray light curve (Fig.~\ref{figure5_modelComparison}b).  These collectively constrain the mass of the shell and the properties of the wind before and after the ejection.

It is difficult to distinguish exact GRB progenitors because many progenitors predict very similar mass ejections.  The morphology and location of the dense shell at the time of the burst are determined by three factors: the delay time between shell ejection and the burst, the mass of the shell, and the wind mass-loss rate of the progenitor.  In general, binary mass ejecta will be slower, but more massive, than stellar eruptions.  Timing alone of a flare, however, does not provide a unique constraint (the position of the shell is a function of its velocity and time between ejection and collapse).  However, some progenitors predict specific structures.  The helium-merger model, for example, predicts a massive shell very close to the exploding star.  The first possible evidence of such a progenitor may be the recent ``Christmas'' burst~\citep{thoneEA11}.  With detailed models, we may be able to place velocity and mass constraints on these shells.  With such information, we will both be able to better understand massive star evolution and constrain the progenitors of GRBs.

In our models we have adopted some approximations and neglected some effects on our light curves.  First, we neglect the spherical nature of the emitting region, which, if included, would undoubtedly result in a modification of our light curves \citep{fenimoreEA96}.  We have investigated this effect, and the dominant result is that the light curve is broadened and becomes more smooth, decreasing the imprint of the circumstellar shell.  For the majority of the duration of the jet's passage through the shell, however, the jet is non-relativistic, which will tend to diminish the effect.  The shell's imprint on the light curve, though lessened, remains detectable, especially at frequencies where the imprint of the shell is largest when spherical emission is ignored.  Reverse shocks might form at the interfaces between regions I and II, regions II and III, and regions IV and V, and could seriously affect the flux of the burst \citep[see i.e.][]{nakarGranot07} at those radii.  Since magnetic fields in GRB jets are not well understood, it is also a simplification to assume that they are in equipartition.  Finally, we present synchrotron light curves only. Inverse Compton scattering may become important after the jet emerges from the dense shell and could increase the flux at high frequencies and late times.  

That said, our semi-analytical method can broadly discriminate between progenitors of GRBs and compute approximate GRB light curves and light curves in general density fields.  Consequently, it is applicable to many other GRB host environments besides uniform media and winds.  Our modeling of the circumburst environment with the Zeus-MP code shows a wide departure from the canonical power law density models, with sudden jumps in density of up to ten orders of magnitude that have a measureable effect on the light curve.  Efforts are now underway to apply this method to model observational signatures and detection thresholds for Population III gamma-ray bursts in primordial H II regions at $z \sim$ 20 \citep{whalenEA04, kitayamaEA04, alvarezEA06, abelEA07, wiseAbel08b} and both Pop II and Pop III GRBs in primeval galaxies at $z \sim$ 10 \citep{wiseEA12} for potential successors to \textit{Swift}, such as the \textit{Joint Astrophysics Nascent Satellite} \citep[\textit{JANUS},][]{roming08,burrowsEA10} and \textit{Lobster}.  

Our method provides a means to obtain a qualitative understanding of gamma-ray burst light curves.  For a more precise calculation, we must turn to computer simulations.  Both special-relativistic magnetohydrodynamical simulations and particle-in-cell (PIC) calculations of GRB jets in circumburst media are now under development, and will soon reveal the light curves of these cosmological explosions in unprecedented detail.

\acknowledgments

RM was supported by LANL IGPP grant 10-150, and DW was supported by the Bruce and Astrid McWilliams Center for Cosmology at Carnegie Mellon University.  Work at LANL was done under the auspices of the National Nuclear Security Administration of the U.S. Department of Energy at Los Alamos National Laboratory under Contract No. DE-AC52-06NA25396.  All ZEUS-MP simulations were performed with allocations from Institutional Computing (IC) on the Conejo cluster at LANL.  

\nocite{alvarezEA06}
\nocite{anninosEA97}
\nocite{abelEA07}
\nocite{barkovKomissarov11}
\nocite{bergerEA00}
\nocite{burrowsEA10}
\nocite{castorEA75}
\nocite{curranEA11}
\nocite{daiLu02}
\nocite{dalgarnoMcCray72}
\nocite{falconeEA06}
\nocite{fenimoreEA96}
\nocite{filgasEA11}
\nocite{frailEA00}
\nocite{fryerEA99b}
\nocite{fryerEA06}
\nocite{fryerEA07b}
\nocite{fryerWoosley98}
\nocite{godetEA06}
\nocite{huangEA99}
\nocite{huangEA00}
\nocite{huangEA12a}
\nocite{kitayamaEA04}
\nocite{meszaros02}
\nocite{mimicaGiannios11}
\nocite{moderskiEA00}
\nocite{nakarGranot07}
\nocite{nardiniEA11}
\nocite{panaitescuKumar00}
\nocite{panaitescuMeszaros00}
\nocite{peer12}
\nocite{priceEA02}
\nocite{ramirez-RuizEA01}
\nocite{ramirez-RuizEA05}
\nocite{ricottiEA01}
\nocite{ricottiEA02}
\nocite{ricottiEA05}
\nocite{roming08}
\nocite{rybickiLightman79}
\nocite{sariEA98}
\nocite{shullvanSteenberg85}
\nocite{thoneEA11}
\nocite{weaverEA77}
\nocite{whalenEA04}
\nocite{whalenEA08b}
\nocite{whalenNorman06}
\nocite{whalenNorman08a}
\nocite{whalenNorman08b}
\nocite{wijersGalama99}
\nocite{wiseAbel08b}
\nocite{wiseEA12}
\nocite{woosley93}
\nocite{woosley11}
\nocite{woosleyBloom06}
\nocite{yostEA03}
\nocite{zhangFryer01}
\nocite{ziaeepourEA08}

\bibliographystyle{apj}
\bibliography{refs}

\end{document}